\documentclass[twocolumn,superscriptaddress,reprint,amsmath,amssymb,aps,prl]{revtex4-1}
\usepackage{graphicx}
\usepackage{dcolumn}
\usepackage{bm}
\usepackage{amsmath}
\usepackage{color}
\usepackage{float}
\usepackage {footnote}

\begin{document}
\title{Generation of high-frequency combs locked to atomic resonances by quantum phase modulation}
\author{Zuoye~Liu}
\affiliation{Max-Planck-Institut f\"ur Kernphysik, Saupfercheckweg 1, 69117 Heidelberg, Germany}%
\affiliation{School of Nuclear Science and Technology, Lanzhou University, 730000, Lanzhou, China}%
\author{ Christian~Ott}
\affiliation{Max-Planck-Institut f\"ur Kernphysik, Saupfercheckweg 1, 69117 Heidelberg, Germany}%
\author{Stefano~M.~Cavaletto}
\affiliation{Max-Planck-Institut f\"ur Kernphysik, Saupfercheckweg 1, 69117 Heidelberg, Germany}%
\author{Zolt$\acute{\text{a}}$n~Harman}
\affiliation{Max-Planck-Institut f\"ur Kernphysik, Saupfercheckweg 1, 69117 Heidelberg, Germany}%
\author{Christoph~H.~Keitel}
\affiliation{Max-Planck-Institut f\"ur Kernphysik, Saupfercheckweg 1, 69117 Heidelberg, Germany}%
\author{Thomas~Pfeifer}
\email[]{tpfeifer@mpi-hd.mpg.de}
\affiliation{Max-Planck-Institut f\"ur Kernphysik, Saupfercheckweg 1, 69117 Heidelberg, Germany}
  \date{\today}
\begin{abstract}
 A general mechanism for the generation of frequency combs referenced to atomic resonances is put forward. The mechanism is based on the periodic phase control of a quantum system's dipole response. We develop an analytic description of the comb spectral structure, depending on both the atomic and the phase-control properties. We further suggest an experimental implementation of our scheme: Generating a frequency comb in the soft-x-ray spectral region, which can be realized with currently available techniques and radiation sources. The universality of this mechanism allows the generalization of frequency-comb technology to arbitrary frequencies, including the hard-x-ray regime by using reference transitions in highly charged ions.
\begin{description}
\item[PACS numbers]
32.80.Qk, 42.50.Gy, 42.62.Eh, 42.62.Fi
\end{description}
\end{abstract}
\maketitle
Precision spectroscopy is undergoing a revolution driven by the invention of frequency combs. These combs represent a ``ruler" of equidistant spectral lines and by now allow to create a fully phase-locked link from radio frequencies through the infrared, visible, all the way up into the extreme-ultraviolet (XUV) spectral region~\cite{Cingoz,Cundiff,Hnsch}. Frequency combs have set off several disruptive developments in various fields, including optical clocks~\cite{Takamoto}, attosecond control of electronic processes~\cite{Baltuska}, precision distance measurement~\cite{Coddington}, astronomical spectroscopy~\cite{Li}, the quantum control of multilevel atomic systems~\cite{Stowe,Barmes}, and the laser cooling of molecules~\cite{Viteau}. There is a strong desire to push combs into higher-frequency regions, e.g. to enable even more precise clockworks helping in the laboratory search for the variation of fundamental constants~\cite{Berengut}.

A broad range of approaches have thus far been used for the generation of frequency combs. Optical frequency combs were first generated by intra-cavity phase modulation~\cite{Kourogi,Ye}. Later, frequency combs were demonstrated that exploit the comb-like frequency structure of mode-locked lasers with stable repetition rate and carrier--envelope phase~\cite{Jones}. This generation scheme was even transferred into the XUV spectral region, utilizing high-order harmonic generation (HHG) (see~\cite{Krausz} for review) within an enhancement cavity~\cite{Gohle,Cingoz}. Characteristic to the current frequency-comb methods is their ``bottom-up" design, by which a comb is created that is referenced to zero frequency, providing an absolute frequency standard.  It appears interesting to explore the idea of whether there could be a complementary ``top-down" approach to frequency combs, e.g., by spanning a comb around a high-energy long-lived x-ray transition, which could then meet the absolute comb somewhere on the way to lower frequencies.

Here, we provide the mechanism for a ``top-down" frequency-comb approach locked to an atomic reference transition. The key idea is conceptually related to sideband generation of continuous-wave (cw) lasers via classical electro-optical modulation~\cite{DeliHaye,Metcalf,Jiang,Wu,Mishra}, dating back to the early roots of frequency combs~\cite{Kourogi,Ye,Murata}. The classical phase modulation is replaced by a quantum-mechanical phase modulation of excited states of atoms, and the cw-laser is replaced by a coherently-excited resonance in any kind of matter (e.g. atoms, M\"ossbauer nuclei, etc.). By locking to an atomic reference, absolute frequencies are not immediately accessible, however relative frequencies are accessible for metrology, with the major advantage of scalability to arbitrary parts of the electromagnetic spectrum. A first example of how the x-ray domain can be accessed with combs was discussed in~\cite{Cavaletto}, however requiring a narrow-band intense x-ray source (not available to date) in combination with a standard optical frequency comb laser. \begin{figure}[!ht]\label{KeyIdeal}
\includegraphics[width=8.4cm]{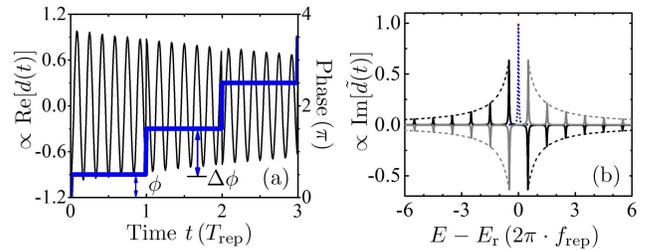}
\caption{(color online) Illustration of the key idea: Phase modulation of the atomic response. (a) Temporal dipole response (thin black line) of a two-level system with periodically applied phase steps (thick blue line) $\Delta\phi=\pi $. The initial phase $\phi$ represents a global offset phase, responsible for line-shape control \cite{Ott} of the comb teeth. (b) Spectral dipole response with the offset phase set to $\phi=\frac{1}{2}\pi$ (black line) and $\phi=\frac{3}{2}\pi$ (grey line): A frequency comb with Lorentzian comb teeth is formed, where the envelope scales as predicted by Eq.~(\ref{ZeroCoefficient}) (dashed lines). For reference, the spectral response of an isolated Lorentzian absorption line is also shown (dotted blue line) with $\Delta\phi=\phi=0$.}
\end{figure}In the mechanism presented in this Letter, we enable the generation of frequency combs using nowadays-available pulsed x-ray sources such as free-electron lasers (FELs)~\cite{Ribic} with a quantum-phase-modulation approach. In this case, the intrinsic natural line width and transition frequency of an atomic resonance defines the resolution and position of the generated comb. Utilizing the results on the time-domain manipulation of spectral line shapes~\cite{Ott}, we demonstrate below how frequency combs can be generated, solely by interaction of available light pulses of duration shorter than the resonance's life time.

The interaction of light with matter can be described via the dipole response function  which is directly proportional to the system's polarizability. For a dilute medium, the imaginary part of this complex-valued function describes absorption or gain of transmitted electromagnetic radiation (see~\cite{Fano, Pollard} for reviews). For a two-level system, i.e.,~an isolated resonance, the time-dependent dipole response $d(t)$ after delta-function-like excitation at time \emph{t}=0 is given by
\begin{equation}\label{TimeDepDipole}
d(t)\propto i e^{-i E_\text{r}t} e^{-\frac{\Gamma}{2}t}\theta(t),
\end{equation}
where $E_{\text{r}}$ and $\Gamma$ denote the energy and the decay width of the excited state, respectively, and $\theta(t)$ is the Heaviside function. Atomic units (a.u.) are used throughout. Through the Fourier transform, the time-dependent dipole response $d(t)$ is related to the frequency-dependent response $\tilde{d}(E)$, $\tilde{d}(E)=\int_{-\infty}^{+\infty}{d(t){{e}^{\text{+}i E t}}\,\mathrm{d}t}$. Computing the imaginary part of the Fourier transform of Eq.~(\ref{TimeDepDipole})
\begin{equation} \label{FrequencyDepDipole}  
\text{Im}\left\{\tilde{d}(E)\right\}\propto\text{Im}\left\{i\frac{1}{\Gamma/2-i\Delta E}\right\}
\end{equation}
yields the well-known Lorentzian line shape, where $\Delta E=E-E_{\mathrm{r}}$ is the energy detuning from the resonance. The decay width $\Gamma$ determines the full width at half maximum (FWHM) of the spectral response, and can generally be interpreted as the coherence decay of the two-level system, subject to spontaneous or other decay mechanisms.

A new phase-control mechanism was demonstrated in~\cite{Ott}, where it was shown how Lorentzian and Fano line shapes can be transformed into each other. This is easily understood by multiplying a phase factor $e^{i\phi}$ to $d(t)$ in Eq.~(\ref{TimeDepDipole}), or to the argument $\tilde{d}(E)$ of the imaginary part in Eq.~(\ref{FrequencyDepDipole}). Experimentally, this is realized by short-pulsed (delta-like) excitation and phase manipulation of the dipole response, which subsequently decays over a much longer time scale, i.e.,~its natural life time $\tau=1/\Gamma$. In general, any phase change $\Delta\phi$ can be realized via impulsive energy shifts $\delta E_{\text{r}}(t)$ and is given by
\begin{equation}\label{GeneralPhaseShift} 
\Delta\phi=\int \delta E_{\text{r}}(t)\textrm{d}t,
\end{equation}
where the integral is to be taken over the duration of the interaction, shifting the energy of the state (e.g.~by using a strong laser pulse). Here we extend this control concept to periodic phase manipulations which cover the whole time of coherence decay of the two-level system.

Fig.~1(a) shows the temporal response of the two-level system after delta-like excitation at \emph{t}=0. Incremental phase steps $\Delta\phi$ are inserted periodically every $ T_{\mathrm{rep}}$, after starting with an initial phase of $\phi$. Via the Fourier transform, the spectral response is revealed: a frequency comb is generated which can be given as an analytical expression without any approximation:
\begin{equation} \label{Comb} 
\begin{split}
\text{Im}\left\{\tilde{d}(E)\right\}&\propto\text{Im}\left\{i e^{i\left(\phi-\Delta\phi/2\right)}\times\right.\\
&\left.\sum\limits_{n=-\infty}^{+\infty}\frac{a_n}{\Gamma/2-i\left[\Delta E-2\pi\cdot{f}_{\mathrm{rep}}\left(n-\Delta {\phi}/2\pi\right)\right]}\right\},
\end{split}
\end{equation}
with comb-teeth coefficients
\begin{equation} \label{ZeroCoefficient} 
a_n=-\frac{\sin(\Delta\phi/2)}{\pi(n-\Delta\phi/2\pi)},
\end{equation}
where \emph{n} is an integer number, and ${f}_{\mathrm{rep}}=\text{1}/{T}_{\mathrm{rep}}$ is the repetition frequency of the applied phase steps $\Delta \phi$. The line shape of each comb tooth is Lorentzian, provided that $\phi=\Delta\phi/2$ as explained below. The global phase $\phi$ equally affects all created comb teeth as it can be taken out of the infinite sum. Eq.~(\ref{Comb}) thus reveals the typical spectral response of a frequency comb spanning across the resonance energy and spaced by integer multiples of the repetition frequency. An illustration of this spectral response is shown in Fig.~1(b). Comparing Eq.~(\ref{Comb}) with Eq.~(\ref{FrequencyDepDipole}), each comb tooth has equal width $\Gamma$ and a common phase factor ${e}^{i(\phi-\Delta\phi/{2})}$, leading to equal line shapes of the comb teeth as set by the global offset phase $\phi$. The strength of each tooth is given by $a_n$ in Eq.~(\ref{ZeroCoefficient}).\begin{figure}[!ht]\label{LocationShift}
\includegraphics[width=6cm]{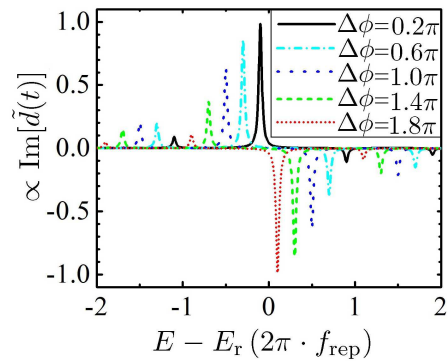}
\caption{(color online) Tuning the teeth of the frequency comb. Using different periodic phase steps $\Delta\phi$, where the offset phase is kept at $\phi=\Delta\phi/2$ to obtain symmetric comb lines, the frequency position of each comb tooth can be tuned continuously over a frequency range $f_{\mathrm{rep}}$.}
\end{figure} As can be seen in Fig.~1(b), symmetric Lorentzian comb teeth are located on each side of the resonance frequency, separated by ${f}_{\mathrm{rep}}$. With parameters set to $\phi =\frac{1}{2}\pi$ and $\Delta \phi =\pi$, the comb with positive energies ($\Delta E>0$) appears with negative amplitudes as gain, while it appears with positive amplitudes as absorption for negative energies. Setting the parameters to $\phi=\frac{3}{2}\pi$ and $\Delta\phi=\pi$, the structure of the created frequency comb is conserved except that absorption and gain sides exchange their roles. The common Lorentzian absorption line shape as described by Eq.~(\ref{FrequencyDepDipole}) is also recovered in Eq.~(\ref{Comb}). If $\phi$ is set to 0 or $\pi$, while keeping $\Delta\phi=0$, a positive (absorption) or negative (gain) Lorentzian line shape results. For the general case $\Delta\phi\ne 0$, comparing with Eq.~(\ref{FrequencyDepDipole}), the offset phase $\phi$ should be set to $\phi =\Delta\phi/2$, or $\phi =\pi+\Delta\phi/2$ in order to keep the Lorentzian symmetric line shape for each comb tooth.

So far, we restricted our discussion to ``maximum" phase steps $\Delta\phi=\pi$. According to Eq.~(\ref{Comb}), another important feature of the impulsive phase-shift method is found: The location of the frequency comb can be finely tuned by setting the incremental phase step $\Delta\phi$ as illustrated in Fig.~2. The location of the \emph{n}-th comb tooth is given by $E_n={E}_{\mathrm{r}}+2\pi\cdot {f}_{\mathrm{rep}}(n-{\Delta\phi}/{2\pi})$, having a symmetric Lorentzian shape with appropriate choice of the initial offset phase (e.g.,~$\phi ={\Delta \phi }/{2}$). The magnitude of the respective comb teeth is also changed as a function of $\Delta \phi$, determined by the coefficients $a_n$ as given by Eq.~(\ref{ZeroCoefficient}). Deviating from the ``maximum" phase step $\Delta\phi=\pi$ will lead to a decrease of the comb-tooth amplitude. For the case of $\Delta\phi =\pi$, 30 comb teeth are produced on each side of the resonance energy $E_{\mathrm{r}}$ above the threshold of 1\% of the original resonance strength, while 3000 comb teeth are generated on each side above the threshold of 0.01\% of the original resonance strength. In the more general case of $\Delta\phi\ne\pi$, the number of comb teeth above a given threshold are reduced by the factor $\sin(\Delta\phi/2)$.
\begin{figure}[!ht]\label{NonzeroComb}
\includegraphics[width=8.4cm]{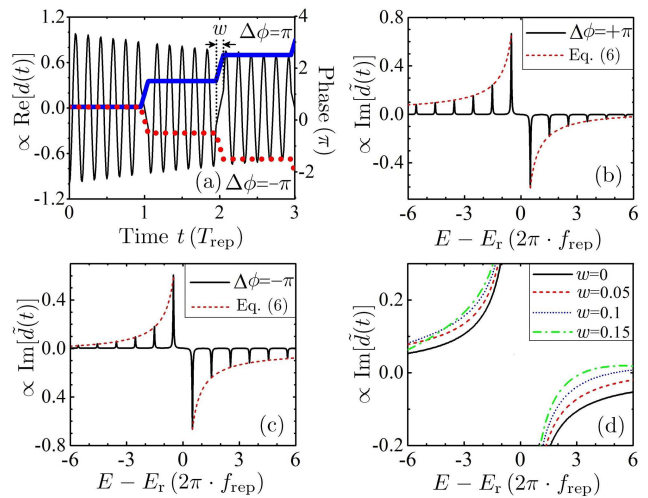}
\caption{(color online) Effects of a non-zero rise time per phase step ($\phi=\pi/2$, $\Delta\phi=\pm\pi$) of duration $w$, in units of the repetition period $T_{\mathrm{rep}}$. (a) Time evolution with positive (thick blue line) and negative (red dotted line) phase steps for a non-zero linear rise time $w$. The oscillating temporal dipole-response function is shown for positive phase steps (thin black line). (b,c) The generated frequency comb with $w=0.05$ for (b) positive phase steps, and (c) negative phase steps. The envelope marking the relative strength of the comb teeth is drawn with a red dotted line. (d) For positive phase steps, the envelope is compared for various values of $w$.}
\end{figure}
 With the thus-far presented formulas [Eqs.~(\ref{Comb}) and (\ref{ZeroCoefficient})], we assumed instantaneous phase shifts $\Delta\phi $ within zero time. To approach realistic situations, in the following we extend our model to finite-duration phase shifts. We assume a linear change of the phase step as illustrated in Fig.~3(a), where the duration $w$ of linear variation is normalized to the repetition period $T_{\mathrm{rep}}$. The spectral response of such a system is again given by Eq.~(\ref{Comb}), but now with a different coefficient
\begin{equation}  \label{NonzeroCoefficient}
a_n=-\frac{\Delta\phi\,\text{sinc}\left\{\pi\left[\left(n-{\Delta\phi}/{2\pi}\right)w+{\Delta\phi}/{2\pi}\right]\right\}}{2\pi(n-{\Delta\phi}/{2\pi})},
\end{equation}
which reduces to Eq.~(\ref{ZeroCoefficient}) for $w=0$. Thus, the regularity of the frequency comb is preserved, except the amplitudes of the comb teeth are now described by Eq.~(\ref{NonzeroCoefficient}). Setting $w=0.05$, frequency combs obtained with positive and negative phase steps are shown in Fig.~3(b) and (c), where the coefficients $a_n$ as given in Eq.~(\ref{NonzeroCoefficient}) are shown as envelope functions. Interestingly, the magnitude of the comb teeth is enhanced on the positive ($\Delta E>0$) and negative ($\Delta E<0$) energy side, respectively [see Fig.~3(b) and (c)], while the ``center of gravity" is shifted into this direction. Note that the direction of the comb teeth (gain/absorption) can still be set by using the offset phase $\phi$ as discussed above. For different rise times $w$ the envelope function of the comb teeth as given by Eq.~(\ref{NonzeroCoefficient}) is plotted in Fig.~3(d). A slight enhancement of selected magnitudes even compared to the ``ideal" zero-width case ($w=0$) is observed.

After the general introduction of the phase modulation of the dipole response for frequency-comb generation, we suggest a first realistic experimental implementation. Here we present theoretical results for the $\text{1s}^\text{2}$--1s\,2p transition in helium or helium-like beryllium in the XUV and soft-x-ray region of the electromagnetic spectrum. Their natural life times are calculated to $\sim$\,$6\,\mathrm{ns}$ (He) and $\sim$\,$8\,\mathrm{ps}$ (Be$^{2+}$). Employing femtosecond pulsed XUV or soft-x-ray radiation from an FEL, the excitation of the resonances can thus be described as delta-like, as required for the phase-modulated response function which we introduced above.

After the excitation, short laser pulses (also in the fs regime) can be utilized to periodically modify the spontaneous decay of the excited $\mathrm{1s\,2p}$ level. This external laser electric field induces periodic AC Stark shifts $\delta E_{\mathrm{r}}(t)$ on the energy of the excited level $E_{\mathrm{r}}$ which are due to the dipole coupling to nearby off-resonant levels such as $\mathrm{1s\,2s}$, $\mathrm{1s\,3s}$, and $\mathrm{1s\,3d}$. The atomic dipole response ${d}(t)$ and its Fourier transform $\tilde{d}(E)$ are obtained with the solution of the master equation~\cite{Scully,Kiffner} which describes the atomic dynamics in the presence of the periodic train of femtosecond pulses in Fig.~4(a). The transition energies are taken from Ref.~\cite{Yerokhin}, whereas decay rates and dipole matrix elements are computed with \texttt{grasp2K}~\cite{Joensson}. The laser field is described classically by the function
\begin{equation}
\mathcal{E}(t) = \cos(\omega_{\mathrm{L}} t)\sum_{j=0}^{\infty}\tilde{\mathcal{E}}(t - t_0 - j T_{\mathrm{rep}}),
\end{equation}
where $t_0$ is the central time of the first pulse, $\omega_{\mathrm{L}}$ is the central frequency, and $\tilde{\mathcal{E}}(t)=\tilde{\mathcal{E}}_{0}\text{sech}(\gamma t)$ is the single-pulse envelope \cite{Rosen, Lewenstein}. Here, the field-strength maximum $\tilde{\mathcal{E}}_{0}$ is associated with the peak intensity $I=c\tilde{\mathcal{E}}_{0}^2/{8\pi}$, and the bandwidth $\gamma$ is related to the FWHM pulse duration $T=2\text{arccosh}(\sqrt{2})/\gamma$ of $|\tilde{\mathcal{E}}(t)|^2$.
We assume near-infrared (NIR) femtosecond laser pulses centered on $1.5 \,\mathrm{eV}$ with 100-fs (FWHM) duration. Each pulse causes a phase shift as shown by Eq.~(\ref{GeneralPhaseShift}), which displays an approximately linear increase with the laser peak intensity, $\Delta\phi\approx -(1/2)\,\alpha \int\tilde{\mathcal{E}}^2(t)\,dt$, with the effective dynamic polarizability $\alpha$ of the $\mathrm{1s\,2p}$ level. The values extracted from our calculations, $\alpha=354\,\text{a.u.}$ for He and $\alpha = -25.7\,\text{a.u.}$ for $\mathrm{Be}^{2+}$, are in good agreement with dynamic polarizabilities calculated for weak constant-amplitude fields~\cite{footnote1,Budker, Drake}. This implies that a phase shift equal to $|\Delta\phi|=\pi$ is obtained by tuning the peak intensity of the train of pulses to $1.37 \times 10^{11} \,\mathrm{W/cm^2}$ and $1.87 \times 10^{12}\,\mathrm{ W/cm^2}$, in the case of He and $\mathrm{Be}^{2+}$, respectively.
\begin{figure}\label{ExperimentalRealization}
\includegraphics[width=8.4cm]{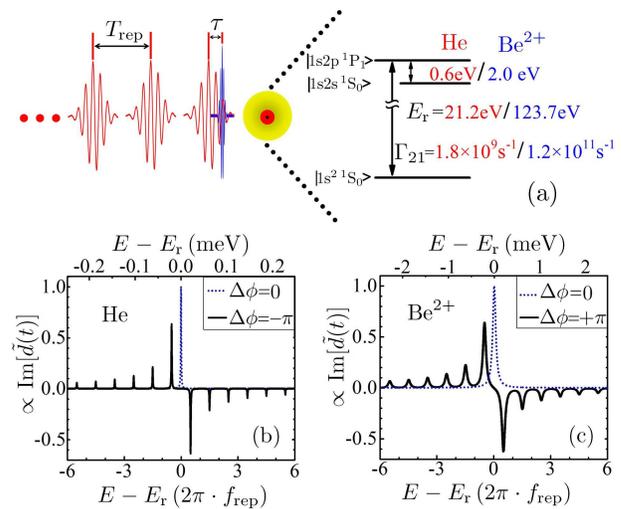}
\caption{(color online) (a) Illustration of a possible experimental realization using pulsed light to create the coherence (blue) and temporally locked pulses (red) for the phase manipulation with the energy level scheme of He (red) and He-like $\text{B}{{\text{e}}^{2+}}$ (blue). In this simplified illustration, only one state (1s2s) is shown, to which the 1s$^2$--1s2p resonance is coupled. The calculated spectral response around the resonant transition is shown (b) for the He atom, and (c) for the $\text{Be}^{2+}$ ion.}
\end{figure}
The time delay $\tau$ between the delta-like excitation and the NIR pulse train holds the key to independently modify the initial (offset) phase $\phi$: It can be set such that the first NIR pulse only partially overlaps with the excitation, thus the phase shift caused by the first NIR pulse (i.e.,~the offset phase) is controlled independently, here tuned to $\phi =\pi/2$. The resulting spectral response of this scheme is shown in Fig.~4(b) and (c), where the repetition frequency of the NIR pulse train is tuned to 10 GHz for He, and to 100 GHz for Be$^{2+}$ in order to separate the individual comb teeth. Both intensity and repetition rate of the NIR pulses are higher in magnitude than typical laser pulses which can be coupled out of a mode-locked oscillator. For a first demonstration of the concept, therefore, we propose to take advantage of amplified Ti:Sa laser pulses coupled into external Fabry--Perot cavities which should easily provide the here assumed specifications. The spectral response of the three-level model shown in Fig.~4(b) and (c) agrees excellently with our previously developed analytical formalism as shown in Fig.~1(b).

In conclusion, frequency combs locked to atomic resonances can be generated by a quantum-phase-modulation method. The two presented examples demonstrate the experimental feasibility of the process in the XUV and soft-x-ray domain. The method itself however can in principle be transferred to any spectral region, including the hard-x-ray domain by using highly charged ion transitions~\cite{Bernitt}. Metastable states will be desirable targets to enhance the comb��s frequency precision. The analytical expressions presented in this work can be readily used to estimate the exact experimental parameters required for the specific system of interest. Though the spectral bandwidth of the here-proposed frequency combs appears limited, we expect a much broader achievable bandwidth by extending the method to more sophisticated phase and amplitude modulation patterns. In all generality the here developed formalism is just one example of the potential provided by arbitrary tailoring of spectral line shapes via direct temporal access to the response function. The generalization of this approach should thus not only enable high-precision metrology and related applications at extremely high frequencies, but also pave the way towards pulse shaping at arbitrary frequencies.

We thank Thomas Allison and Itan Barmes for a critical reading of the manuscript and providing very helpful comments. The authors acknowledge support from the Max-Planck Research Group program and DFG (grant PF790/1-1). Z.~Liu acknowledges the scholarship award for excellent doctoral student granted by the Ministry of Education of China.


\begin{thebibliography}{99}
  \bibitem{Cingoz}
   A.~Cing\"oz, D.~C.~Yost, T.~K.~Allison, A.~Ruehl, M.~E.~Fermann, I.~Hartl, and J.~Ye, Nature \textbf{482}, 68 (2012).

  \bibitem{Cundiff}
   S.~T.~Cundiff and J.~Ye, Rev. Mod. Phys. \textbf{75}, 325 (2003).

  \bibitem{Hnsch}
   T.~W.~H\"ansch, Rev. Mod. Phys. \textbf{78}, 1297 (2006).

  \bibitem{Takamoto}
   M.~Takamoto, F.~L.~Hong, R.~Higashi, and H.~Katori, Nature \textbf{435}, 321 (2005).

  \bibitem{Baltuska}
   A.~Baltu$\check{\text{s}}$ka, T.~Udem, M.~Uiberacker, M.~Hentschel, E.~Goulielmakis, C.~Gohle, R.~Holzwarth, V.~S.~Yakovlev, A.~Scrinzi, T.~W.~H\"ansch, and F.~Krausz, Nature \textbf{421}, 611 (2003).

  \bibitem{Coddington}
   I.~Coddington, W.~C.~Swann, L.~Nenadovic, and N.~R.~Newbury, Nat. Photonics \textbf{3}, 351 (2009).

  \bibitem{Li}
   C.~H.~Li, A.~J.~Benedick, P.~Fendel, A.~G.~Glenday, F.~X.~K\"artner, D.~F.~Phillips, D.~Sasselov, A.~Szentgyorgyi, and R.~L.~Walsworth, Nature \textbf{452}, 610 (2008).

  \bibitem{Stowe}
   M.~C.~Stowe, A.~Pe'er, and J.~Ye, Phys. Rev. Lett. \textbf{100}, 203001 (2008).

   \bibitem{Barmes}
   I.~Barmes, S.~Witte, and K.~S.~E.~Eikema, Nat. Photonics \textbf{7}, 38 (2013).

  \bibitem{Viteau}
   M.~Viteau, A.~Chotia, M.~Allegrini, N.~Bouloufa, O.~Dulieu, D.~Comparat, and P.~Pillet, Science \textbf{321}, 232 (2008).

   \bibitem{Berengut}
   J.~C.~Berengut, V.~A.~Dzuba, V.~V.~Flambaum, and A.~Ong, Phys. Rev. Lett. \textbf{106}, 210802 (2011).

  \bibitem{Kourogi}
   M.~Kourogi, K.~Nakagawa, and M.~Ohtsu, IEEE J. Quantum Electron. \textbf{29}, 2693 (1993); M.~Kourogi, T.~Enami, and M.~Ohtsu, IEEE Photonic Technol. Lett. \textbf{6}, 214 (1994).

  \bibitem{Ye}
   J.~Ye, L.~Ma, T.~Daly, and J.~L.~Hall, Opt. Lett. \textbf{22}, 301 (1997).

  \bibitem{Jones}
   D.~J.~Jones, S.~A.~Diddams, J.~K.~Ranka, A.~Stentz, R.~S.~Windeler, J.~L.~Hall, and S.~T.~Cundiff, Science \textbf{288}, 635 (2000).

    \bibitem{Krausz}
   F.~Krausz and M.~Ivanov, Rev. Mod. Phys. \textbf{81}, 163 (2009).

   \bibitem{Gohle}
   C.~Gohle, T.~Udem, M.~Herrmann, J.~Rauschenberger, R.~Rolzwarth, H.~A.~Sch\"ussler, F.~Krausz, and T.~W.~H\"ansch, Nature \textbf{436}, 234 (2005).

  \bibitem{Cavaletto}
   S.~M.~Cavaletto, Z.~Harman, C.~Buth, and C.~H.~Keitel, arXiv:1302.3141.

  \bibitem{DeliHaye}
   P.~Deli'Haye, A.~Schliesser, O.~Arcizet, T.~Wilken, R.~Holzwarth, and T.~J.~Kippenberg, Nature \textbf{450}, 1214 (2007).

  \bibitem{Metcalf}
   A.~J.~Metcalf, D.~E.~Leaird, and A.~M.~Weiner, IEEE J. Sel. Top.  Quantum Electron. \textbf{19}, 3500306 (2103).

  \bibitem{Jiang}
   Z.~Jiang, C.~Huang, D.~E.~Leird, and A.~M.~Weiner Nat. Photonics \textbf{1}, 463 (2007).

  \bibitem{Wu}
   R.~Wu, V.~R.~Supradeepa, C.~M.~Long, D.~E.~Leaird, and A.~M.~Weiner, Opt. Lett. \textbf{35}, 3234 (2010).

  \bibitem{Mishra}
   A.~K.~Mishra, R.~Schmogrow, I.~Tomkos, D.~Hillerkuss, C.~Koos, W.~Freude, and J.~Leuthold, IEEE Photonic Technol. Lett. \textbf{25}, 701 (2013).

  \bibitem{Murata}
   H.~Murata, A.~Morimoto, T.~Kobayashi, and S.~Yamamoto, IEEE J. Sel. Top.  Quantum Electron. \textbf{6}, 1325 (2000).

     \bibitem{Ribic}
  P.~R.~Ribic and G.~Margaritondo, J. Phys. D: Appl. Phys. 45 213001 (2012).

  \bibitem{Ott}
   C.~Ott, A.~Kaldun, P.~Raith, K.~Meyer, M.~Laux, J.~Evers, C.~H.~Keitel, C.~H.~Greene, and T.~Pfeifer, Science \textbf{340}, 716 (2013).

  \bibitem{Fano}
   U.~Fano and J.~W.~Cooper, Rev. Mod. Phys. \textbf{40}, 441 (1968).

  \bibitem{Pollard}
   W.~T.~Pollard and R.~A.~Mathies, Annu. Rev. Phys. Chem. \textbf{43}, 497 (1992).

   \bibitem{Scully}
   M.~O.~Scully and M.~S.~Zubairy, \textit{Quantum Optics} (Cambridge University Press, Cambridge, 1997).

      \bibitem{Kiffner}
      M.~Kiffner, M.~Macovei, J.~Evers, and C.~H.~Keitel, ``\textit{Vacuum-induced processes in multi-level atoms}" in ``\textit{Progress in Optics}" E. Wolf, ed. (Elsevier, Amsterdam, 2010), pp.8--197.


    \bibitem{Yerokhin}
     V.~A.~Yerokhin and K.~Pachucki, Phys. Rev. A \textbf{81}, 022507 (2010).

\bibitem{Joensson}
P.~J\"onsson, X.~He, C.~F.~Fischer, and I.~P.~Grant, Comput. Phys. Commun. \textbf{177}, 597 (2007).

\bibitem{Rosen}
 N.~Rosen and C.~Zener, Phys. Rev. \textbf{40}, 502 (1932).

\bibitem{Lewenstein}
  M.~Lewenstein, J.~Zakrzewski, and K.~Rzazewski, J. Opt. Soc. Am. B \textbf{3}, 22 (1986).

\bibitem{footnote1}
These values neglect the small contributions due to the Bloch--Siegert shift~\cite{Budker}.

\bibitem{Budker}
  D.~Budker, D.~F.~Kimball, and D.~P.~DeMille, \textit{Atomic Physics}, (Oxford University Press, Oxford, 2004).

\bibitem{Drake}
G.~W.~F.~Drake, ``\textit{High precision calculations for helium},'' in ``\textit{Springer Handbook of Atomic,
Molecular, and Optical Physics},'' G.~W.~F.~Drake, ed. (Springer, New York, 2006), pp. 199--219.

    \bibitem{Bernitt}
   S.~Bernitt \emph{et al.}, Nature \textbf{492}, 225 (2012).

\end{thebibliography}
\end{document}